\documentclass[prd,aps,secnumarabic,superscriptaddress,floatfix,preprintnumbers]{revtex4-1}
\usepackage[dvips]{graphicx,color}
\usepackage{dcolumn}
\usepackage{amsmath}
\usepackage{rotating}
\def\mc{\multicolumn}
\def\sa{\mbox{\rule{0pt}{10pt}}}

\def\st{\mbox{\rule[-3pt]{0pt}{14pt}}}
\def\stt{\mbox{\rule[-4pt]{0pt}{16pt}}}

\def\Dot{\!\cdot\!}
\def\al{\alpha}
\def\be{\beta}
\def\ga{\gamma}
\def\de{\delta}
\def\De{\Delta}

\def\la{\lambda}

\def\ro{\rho}

\def\ep{\varepsilon}
\def\part{\partial}
\def\arccosh{\mbox{\rm arccosh}}

\def\Li{{\rm Li}}
\def\Re{\mbox{\rm Re}}

\newcommand{\mathsym}[1]{{}}
\newcommand{\unicode}[1]{{}}

\begin{document}

\title{Vacuum polarization corrections to $\mu^+\mu^-$ spectrum}
\author{Wayne W. Repko}\email{repko@pa.msu.edu}
\affiliation{Department of Physics and Astronomy, Michigan State University, East Lansing, MI 48824}
\date{\today}
\begin{abstract}
The energy levels of the $n=1$ and $n=2$ bound states of the $\mu^+\mu^-$ atom (true muonium) are calculated starting from a previously derived potential that correctly describes positronium to order $\al^5$. All electron vacuum polarization corrections on the true muonium levels are computed to the same $\al^5$ order and a few of order $\al^6$ are examined to get a sense of size of these contributions. Additional order $\al^6$ contributions from the two photon and three photon annihilation processes are included in the evaluation of the ground state ($1^3S_1-1^1S_0$) hyperfine splitting.
\end{abstract}

\maketitle

\section{Introduction}

Since the discussion of the $\mu^+\mu^-$ bound system (true muonium) by Hughes and Maglic \cite{HM} in 1971 and a calculation of the ground state hyperfine splitting due to electron vacuum polarization shortly thereafter \cite{OR}, the topic remained relatively dormant until the late 1990's \cite{JSIK1997}. Recently, with the increasing prospect of experimental searches for these bound states, there has been a renewed theoretical interest in the $\mu^+\mu^-$ atom \cite{Lamm:2013oga,Lamm:2015lia,Lamm:2016djr,Ji:2016fat,Lamm:2017lib} and numerous discussions about the best method to discover true muonium \cite{BL,Banburski:2012tk,Itahashi:2015fra,Sakimoto:2015vrf,Ji:2017lyh}. The calculations that follow amount to another take on the spectrum of the $\mu^+\mu^-$ bound system. They start from a non-relativistic positronium potential that contains all relativistic and QED one-loop corrections. As usual, this potential is supplemented by numerous electron vacuum polarization contributions to the $\mu^+\mu^-$ spectrum. Brief sketches of the electron vacuum polarization calculations are given below and the individual results are contained in Table \ref{T1}. A more detailed description of these calculations can be found in \cite{DR}.  The net result may add something to the theoretical discussion of true muonium particularly in the values of the $2P\to 2S$ transitions.

\section{Vacuum Polarization Corrections}

A convenient starting point for calculation of the true muonium energy level spectrum is the perturbative potential that describes the spectrum of positronium. A potential with this property is derived in Ref.\cite{GRS} and includes all QED corrections to order $\al^5$. It consists of a direct potential $V'(\vec{r})$ and an annihilation potential $V''(\vec{r})$ given by
\begin{eqnarray} \label{VDIR}
V'(\vec{r})&=& -\al\left[\frac{1}{r}+\frac{1}{m^2r}\vec{p}^{\,2} -\frac{4\pi}{3m^2}\vec{S}^{\,2}\de(\vec{r})-\frac{1}{2m^2r^3}S_{12} -\frac{3}{2m^2r^3}\vec{L}\Dot\vec{S}\,\right] \nonumber \\ [6pt]
&&\frac{\al^2}{m^2}\left[\frac{m}{4r^2}+\left(\frac{26}{15}-\frac{14}{3}\ln(2) +\frac{16}{3}\ln(m/\la)\right)\de(\vec{r}) +\frac{7}{6\pi}\nabla^2\left(\frac{\ln(mr/2)+\ga}{r}\right)\right.\nonumber \\ [6pt]
&&\left.\hspace{25pt}-\frac{2}{3}\vec{S}^{\,2}+\frac{1}{2\pi r^3}S_{12}+ \frac{1}{\pi r^3}\vec{L}\Dot\vec{S}\,\right]\,,
\end{eqnarray}
and
\begin{equation} \label{VANN}
V''(\vec{r})=\frac{\al\pi}{m^2}\vec{S}^{\,2}\de(\vec{r}) -\frac{\al^2}{m^2}\left[\left(\frac{26}{9}+ 2\ln(2)\right)\vec{S}^{\,2}+4\left(1-\ln(2)\right)\right]\de(\vec{r})\,,
\end{equation}
where $\vec{S}$ is the total spin, $\ga$ is Euler's constant, $S_{12}=3\vec{S}\Dot\hat{r}\vec{S}\Dot\hat{r}-\vec{S}^{\,2}$ and $\la$ is the infrared cutoff that is removed using the Bethe sum. The $\al^4$ and $\al^5$
contributions to the energy levels from these terms are denoted by $E_2$ and $E_4$ in Table \ref{T1}.
\begin{sidewaystable} 
\begin{tabular}{|l|r|r|r|r|r|r|r|r|r|r|r|r|r|r|r|r|r|}
\hline
\stt & \mc{1}{c|}{$E_2$}& \mc{1}{c|}{$E_4$}& \mc{1}{c|}{$E^{(2)}_{VP}$} & \mc{1}{c|}{$E^{(4)}_{VP}$}& \mc{1}{c|}{$E^{(6)}_{VP}$} & \mc{1}{c|}{$E_A$}& \mc{1}{c|}{$E_{A\,KS}$}&\mc{1}{c|}{$E_{VP\,VP}$}& \mc{1}{c|}{$E_{VP\,KS}$}& \mc{1}{|c|}{$\De\,HF$}& \mc{1}{|c|}{$\De\,LS$}  &\mc{1}{|c|}{$\De\,TEN$} &\mc{1}{|c|}{$\De\,KE$} &\mc{1}{|c|}{$\De\,p^2/r$} &\mc{1}{|c|}{$\De\,\al^2/r^2$} &\mc{1}{|c|}{$E_{\rm Tot}$} &\mc{1}{|c|}{$E_0$}\\
\hline
\st $1^3S_1$ &76.4645 &1.48205 &-485.151 &-7.6999 & -0.0043 &0.58404 &0.006013 &-0.32260 &-0.00489 &0.38576  & & &-0.03172 &-0.25414 &0.04811 &-414.4937 &-1.4166\,10$^6$ \\
\hline
\st $1^1S_0$ &-98.3116 &2.34188 &-485.151 &-7.6999 &-0.0043 & & &-0.32260 &-0.00489 & & & &-0.03172 &-0.25414 &0.04811 &-589.3858 &-1.4066\,10$^6$ \\
\hline
\st $2^3P_2$ &-0.83877 &0.00077 &-1.4297  &-0.01530 &0.00002 & & &-0.00009 &-0.000036 &-0.00008 &0.00037 &-0.000037 &-0.00003 &-0.00018 &-0.00006 &-2.2831 &-3.5165\,10$^5$ \\ 
\hline
\st $2^3P_1$ &-4.5840 &-0.00430 &-1.4297 &-0.01530 &0.00002 & & &-0.00009  &-0.000036 &-0.00008 &-0.00037 &0.00019 &-0.00003 &-0.00018 &-0.00006 &-6.0339 & -3.5165\,10$^5$ \\
\hline
\st $2^3P_0$ &-9.26547 &-0.01336 &-1.4297 &-0.01530 &0.00002 & & &-0.00009 &-0.000036 &-0.00008 &-0.00073 &-0.00037 &-0.00003 &-0.00018 &-0.00006 &-10.7254 &-3.5165\,10$^5$ \\
\hline
\st $2^1P_1$ &-3.02347 &-0.00249 &-1.4297 &-0.01530 &0.00002 & & &-0.00009 &-0.000036 & & & &-0.00003 &-0.00018 &-0.00006 &-4.47133 & -3.5165\,10$^5$\\
\hline
\st $2^3S_1$ &6.33953 &0.19814 &-58.2401 &-0.45736 & -0.0006 &0.07754 &0.000752 &-0.02917 &-0.00043 &0.044566 & & &-0.00327 &-0.02795 &0.00506 &-52.0927 &-3.5165\,10$^5$ \\
\hline
\st $2^1S_0$ &-15.5075 &0.30562 &-58.2401 &-0.45736 &-0.0006  & & &-0.02917 &-0.00043 & & & &-0.00327 &-0.02795 &0.00506 &-73.9551 &-3.5165\,10$^5$ \\
\hline
\end{tabular}
\caption{The entries summarize the various corrections (in meV) to the $n=1$ and $n=2$ states of true muonium. $E^{(2)}_{VP}$ includes the pion one-loop contribution. \label{T1}}
\end{sidewaystable}

The large electron vacuum polarization correction to the Coulomb interaction can be included by using the dispersion representation for the photon propagator \cite{GK,HL,KS}
\begin{equation}\label{prop}
D(k^2)=-\frac{1}{k^2}+\int_0^\infty\!\!\frac{d\la}{\la}\frac{\Delta(\la)}{\la-k^2-i\ep}\,,
\end{equation}
where $\Delta(q^2)$ is
\begin{equation}
\Delta(q^2)=-\frac{(2\pi)^3}{3q^2}\sum_n\de^{(4)}(q-q_n)\langle0|j_\mu(0)|n\rangle \langle n|j^\mu(0)|0\rangle\,.
\end{equation}
For the $e\,\bar{e}$ intermediate state, the leading contribution, $\Delta^{(2)}(\la)$, is
\begin{equation}
\Delta^{(2)}(\la)=\frac{\al}{3\pi}(1+2m_e^2/\la)\sqrt{1-4m_e^2/\la}\;\theta(\la-4m_e^2)\,.
\end{equation}
If we take $k^2$ to be space-like, then the modified Coulomb interaction is
\begin{eqnarray}\label{spacelike}
V(\vec{k}^{\,2}) &=&-\frac{e^2}{\vec{k}^{\,2}}-e^2\!\!\int_{4m_e^2}^\infty\!\frac{d\la}{\la} \frac{\Delta^{(2)}(\la)}{\vec{k}^{\,2}+\la} \\
&=& V_C(\vec{k}^{\,2})+V_{VP}(\vec{k}^{\,2})\,.
\end{eqnarray}
Transforming to coordinate space
\begin{eqnarray}
V_{VP}(\vec{r})&=&\frac{1}{(2\pi)^3}\int\!\!d^{\,3}k\,V_{VP}(\vec{k})e^{i\vec{k}\cdot\vec{r}}\\ [6pt]
V_{VP}(\vec{r})&=&-\frac{\al}{r} \int_{4m_e^2}^\infty\!\frac{d\la} {\la}\Delta^{(2)}(\la)e^{-\sqrt{\la}\,r}\,.
\end{eqnarray}
The first order spin-independent contributions of $V_{VP}(\vec{r})$ to the $1S$, $2S$ and $2P$ states are given in the $E^{(2)}_{VP}$ column of Table \ref{T1}. For example, the leading $e^+e^-$ vacuum polarization correction to the ground state is
\begin{eqnarray}
\langle 1S|V_{VP}|1S\rangle &=& \int_0^\infty\!\!dr\,r^2R_{10}^{\,2}(r)\,V_{VP}(r) \nonumber \\
&=& -\frac{2m_\mu\,\al^3}{3\pi}\int_4^\infty\!\! \frac{dx}{x}\frac{(1+2/x)\sqrt{1-4/x}}{(2+\be\sqrt{x})^2}=-485.143\,{\rm meV}\,,\label{VP}
\end{eqnarray}
where $\be=2m_e/m_\mu\al$. 

There is also a fourth order contribution to $\Delta(\la)$, $\Delta^{(4)}(\la)$, due to K\"allen and Sabry \cite{KS}. It results in a potential $V_{KS}(r)$ whose contributions are found in the $E^{(4)}_{VP}$ column of Table \ref{T1}. The ground state contribution has the form that resembles Eq.\,(\ref{VP})
\begin{equation}
E^{(4)}_{VP}(1S)=\frac{2m_\mu\,\al^4}{\pi^2}\int_4^\infty\!\!\frac{dx}{x} \frac{\Delta^{(4)}(\sqrt{x}/2)}{(2+\be\sqrt{x})^2}=-7.6999\,{\rm meV}\,,
\end{equation}
with
\begin{eqnarray}
\Delta^{(4)}(x) &=& \left(\frac{13}{54 x}+\frac{7}{108 x^3}+\frac{2}{9 x^5}\right) \sqrt{x^2-1}+\left(\frac{4}{3 x}+\frac{2}{3 x^3}\right) \sqrt{x^2-1} \log\left(8x \left(x^2-1\right)\right) \nonumber\\ [6pt]
&&+\left(-\frac{44}{9}+\frac{2}{3 x^2}+\frac{5}{4 x^4}+\frac{2}{9 x^6}\right) \arccosh(x)+\left(-\frac{8}{3}+\frac{2}{3
x^4}\right)\Bigg[\frac{2 \pi ^2}{3} \nonumber\\ [6pt]
&&-\arccosh^2(x)-\log\left(8x(x^2-1)\right)\arccosh(x)-2\Re[\Li_2\left((x+\sqrt{x^2-1})^2\right)] \nonumber \\ [2pt]
&&+\Li_2\left(-(x-\sqrt{x^2-1})^2\right)\Bigg]\,.
\end{eqnarray}

In addition, there is a sixth order contribution to $\Delta(\la)$, $\Delta^{(6)}(\la)$, which contributes to order $\al^5$. The precise analytical form $\Delta^{(6)}(\la)$ is not known. It is comprised of a pair of one-particle reducible terms: one containing three electron loops and the other consisting of an electron loop connected to an electron loop that has one internal photon line. These contributions are accompanied by an irreducible three loop bubble diagram \cite{KN}. The reducible diagrams can be calculated using $\Delta^{(2)}(\la)$ and $\Delta^{(4)}(\la)$. The calculation of the irreducible diagram is not included here. These corrections appear in Table \ref{T1} as $E^{(6)}_{VP}$.

The potential $V_{VP}(\vec{r})$ can also contribute in the second order of non-relativistic perturbation theory and these contributions are found in the $E_{VP\,VP}$ column in Table \ref{T1}. For the ground state, $E_{VP\,VP}(1S)$ is obtained from
\begin{equation} \label{VPVP}
E_{VP\,VP}(1S)=\frac{2m_\mu\,\al^4}{9\pi^2}\int_4^\infty\!\!\frac{d\la}{\la}\Pi^{(2)}_e(\la) \int_4^\infty\!\!\frac{d\ro}{\ro}\Pi^{(2)}_e(\ro)\int_0^\infty\!\!dx\,xe^{-\be\sqrt{\la}x}e^{-x} \int_0^\infty\!\!dx'x'g_{10}(x,x')e^{-\be\sqrt{\ro}x'}e^{-x'}\,,
\end{equation}
where $g_{10}(x,x')$ is the Coulomb radial Green's function
\begin{equation}
g_{10}(x,x')=4e^{-x}e^{-x'}\left[\ln(2x)+\ln(2x')+x+x'+2\ga-\frac{7}{2}-\frac{1}{2x} -\frac{1}{2x'}+\frac{e^{2x_<}}{2x_<}-Ei(2x_<)\right]\,.
\end{equation}
Here $Ei(x)$ is the exponential integral and $x_<$ is the smaller of $x$ and $x'$. The $x$ and $x'$ integrals can be evaluated analytically using Mathematica and the remaining integrals over $\Delta^{(2)}(\la)$, when evaluated using NIntegrate, give $-0.04782$. With this, $E_{VP\,VP}(1S)=-0.32260$ meV.

In addition, there are order $\al^5$  contributions from the combination of $V_{VP}(r)$ and $V_{KS}(r)$ and these are given in the $E_{VP\,VK}$ column of Table \ref{T1}. The calculation analogous to Eq.\,(\ref{VPVP}) with $\Delta^{(4)}(\sqrt{\ro}/2)$ replacing $\Delta^{(2)}(\ro)$ together with adding a factor of 2 and an adjustment in the couplings. There are also order $\al^5$ contributions from third order non-relativistic perturbation theory involving three $V_{VP}$ potentials, but these are negligible.

The electron vacuum polarizations computed thus far are spin-independent. Since the $\mu^+\mu^-$ bound system \cite{HM}  has an annihilation channel, there are corrections associated with this process. Some of them are contained in Eq.\,(\ref{VANN}), but the largest effect is the electron vacuum polarization correction, which depends on $\vec{S}^{\,2}$ \cite{OR}. The energy associated with the annihilation process is
\begin{equation}
E_A(nS)=-\frac{m_\mu\,\al^4}{8n^3}\langle\vec{S}^{\,2}\rangle\, 4m_\mu^2 \int_{4m_e^2}^\infty\!\!\frac{d\la}{\la}\frac{\Delta(\la)}{(\la-4m_\mu^2-i\ep)}\,.
\end{equation}
For $\Delta^{(2)}(\la)$, the integral can be evaluated analytically \cite{OR} and the result is $-3.4609\,\al/\pi$, leading to the entries in the $E_A$ column of Table \ref{T1}. The analogous calculation using $\Delta^{(4)}(\la)$ was performed numerically and those results are in the $E_{A\;KS}$ column of Table \ref{T1}.

The columns in Table \ref{T1} labelled with a $\De$ contain the electron vacuum polarization corrections to the terms in Eq.\,(\ref{VDIR}) that contribute to order $\al^4$ as well as the second order contribution from the kinetic energy term $-(\vec{p}^{\,2})^2/4m_\mu^3$. The second order perturbative corrections are a straight forward calculation of $V_{VP}(r)$ with one of the terms in Eq.\,(\ref{VDIR}) using the appropriate Green's functions \cite{DR}. To illustrate the direct inclusion of the vacuum polarization effect, consider the hyperfine term $4\pi\al\vec{S}^{\,2}\de(\vec{r})/(3m_\mu^2)$. This term arises from the momentum space expression
\begin{equation}
e^2\frac{1}{3m_\mu^2}\vec{S}^{\,2}=\frac{e^2}{\vec{k}^{\,2}}\, \frac{\vec{k}^{\,2}}{3m_\mu^2}\vec{S}^{\,2}\rightarrow \frac{e^2}{3m_\mu^2}\,\int_{4m_e^2}^\infty\!\frac{d\la}{\la}\,\Delta^{(2)}(\la) \,\frac{\vec{k}^{\,2}}{\vec{k}^{\,2}+\la}\,\vec{S}^{\,2}\,.
\end{equation}
In other words, from Eq.\,(\ref{spacelike}), the vacuum polarization correction is obtained by replacing $-e^2/\vec{k}^{\,2}$ by $V_{VP}(\vec{k}^{\,2})$. One can then transform to coordinate space and calculate the relevant expectation values. The procedure of replacing $-e^2/\vec{k}^{\,2}$ by $V_{VP}(\vec{k}^{\,2})$ works for all of the order $\al^4$ terms.

The final two columns in Table \ref{T1} are $E_{\rm Tot}= E_2+E_4 + {\rm electron\;vacuum\;polarization \;modifications}$ and $E_0$, the unperturbed energy.

The pion vacuum polarization correction is included in $E_{VP}^{(2)}$ and $E_A$. In addition, the $\al^6\log(\al^{-1})$ ground state hyperfine correction as well as the $\al^6$ modified singlet annihilation contribution \cite{AAB,KIJS} and the $\al^6$ triplet annihilation contribution \cite{Cung,Cung1,ABZ} are included in $\De(1^3S_1\to 1^1S_0)$. The evaluation of these terms is contained in the Appendix. The frequencies of various transitions are shown in Table \ref{T2}. These results are compared with Ref.\,\cite{Lamm:2017}.
\begin{table}[h!]\center
\begin{tabular}{|c|r|r|}
\hline
\mc{1}{|c|}{\st Transition} & \mc{1}{c|}{Theory\,(MHz)}& \mc{1}{c|}{Ref.\,(20)} \\
\hline
\st $\De$(1$^3$S$_1$$\to$1$^1$S$_0$) & 4.229297\,10$^7\;^\dag$ & 4.2328355(51)\,10$^7$\\
\hline
\st $\De$(2$^3$S$_1$$\to$2$^1$S$_0$) & 5.28631\,10$^6$ &  \\
\hline
\st $\De$(2$^3$S$_1$$\to$1$^3$S$_1$) & 2.55173\,10$^{11}$ & 2.550014(16)\,10$^{11}$ \\
\hline
\st $\De$(2$^3$P$_2$$\to$2$^3$S$_1$) & 1.20440\,10$^7$ & 1.206(3)\,10$^7$  \\
\hline
\st $\De$(2$^3$P$_1$$\to$2$^3$S$_1$) & 1.11371\,10$^7$ & 1.115(3)\,10$^7$ \\
\hline
\st $\De$(2$^3$P$_0$$\to$2$^3$S$_1$) & 1.00027\,10$^7$ & 1.002(3)\,10$^7$ \\
\hline
\st $\De$(2$^1$P$_1$$\to$2$^3$S$_1$) & 1.15149\,10$^7$ & 1.153(3)\,10$^7$ \\
\hline
\end{tabular}
\caption{The transition frequencies between various levels are shown. Included are the $\pi^+\pi^-$ vacuum polarization corrections. The $^\dag$ indicates that the $\al^6\log(\al^{-1})$ hyperfine correction ($3.95454\,10^3\;{\rm MHz}$), the $\al^6$  (modified) singlet annihilation contribution \cite{AAB,KIJS} ($-285.18\;{\rm MHz}$) and the $\al^6$ triplet annihilation contribution \cite{Cung,Cung1,ABZ} ($-200.38\;{\rm MHz}$) to the ground state are included. The results are compared with those of Ref.\,\cite{Lamm:2017}. \label{T2}}
\end{table} 

\section{Conclusions}

Comparing the results in Table \ref{T2}, there are differences in the ground state hyperfine splitting and the $2^3S_1-1^3S_1$ splitting that are outside the quoted errors. On the other hand, the $2P-2S$ frequencies are consistent those in Ref.\,\cite{Lamm:2017}. To assess the errors from uncalculated order $\al^6$ terms, three $\al^6$ were calculated: the second order $V_{KS}$ correction, the $\al^6$ hyperfine contribution from Eq.\,(\ref{VDIR}) and the annihilation contribution of the K\"allen-Sabry term. The first is completely negligible and second contributes $-0.0125\;10^3$ MHz to the $2^3S_1$ state. The third, which is included in Table \ref{T1}, amounts to a change of $0.183\;10^3$ MHz, in the $2P-2S$ splitting, which suggests a conservative estimate of the uncertainty in $\De(2^3P_2\to2^3S_1)=1.20439(5)\;10^7$ MHz. 

\acknowledgements
The author would like to thank Stan Brodsky and Richard Lebed for conversations and email communications regarding this work.
\appendix
\section{Ground state hyperfine splitting corrections}
In order to generalize the positronium result for the $\al^6$ two photon annihilation correction, given in the paper of G.~S.~Adkins, Y.~Aksu and M.~H.~T.~Bui \cite{AAB}, to the $\mu^+\mu^-$ atom one can simply scale the vertex, self energy and box contributions by replacing the electron mass with the muon mass. However, the vacuum polarization contribution contains an additional contribution due to the electron in the loop. To include this contribution, the calculation in Ref. \cite{AAB} must be generalized.

According to Ref. \cite{AAB}, the vacuum polarization correction for $\mu^+\mu^-$ can be expressed as
\begin{equation}
\Delta E_{VP}=\frac{2m_\mu^3\al^5}{\pi}\!\int\!\frac{d^{\,4}k}{i\pi^2} \frac{\vec{k}^{\,2}\,\Pi_f(k^2)}{k^2\,(k-K)^2\,(k^2-k\Dot K)^2}\,,
\end{equation}
where $K=(2m_\mu,\vec{0})$ and $\Pi_f(k^2)$ is the renormalized vacuum polarization factor with an arbitrary mass in the loop. To proceed, we can use the dispersion representation
\begin{equation}
\frac{\Pi_f(k^2)}{k^2}=\int_{4m^2}^\infty\!\frac{d\la}{\la}\frac{\De (\la)}{\la-k^2-i\ep}\,.
\end{equation}
For the leading order, $\De (\la)$ is
\begin{equation}
\De (\la)=\frac{\al}{3\pi}(1+\frac{2m^2}{\la})\sqrt{1-\frac{4m^2}{\la}}\,\theta (\la-4m^2).
\end{equation}
Expressing $\vec{k}^{\,2}$ as
\begin{equation}
\vec{k}^{\,2}=\frac{(k\Dot K)^2}{K^2}-k^2\,,
\end{equation}
$\De E_{VP}$ is
\begin{equation}
\De E_{VP}=-\frac{2m_\mu^3\al^5}{\pi}\!\int_{4m^2}^\infty\!\frac{d\la}{\la}\De (\la) \int\!\frac{d^{\,4}k}{i\pi^2} \frac{(k\Dot K)^2/K^2-k^2}{(k-K)^2\,(k^2-\la)\,(k^2-k\Dot K)^2}\,.
\end{equation}
Shifting $k$ to $k+K$ and expressing $k$ is units of $m_\mu$ gives
\begin{equation}
E_{VP}=-\frac{2m_\mu\al^5}{\pi}\!\int_{4m^2}^\infty\!\frac{d\la}{\la}\De (\la) \int\!\frac{d^{\,4}k}{i\pi^2} \frac{(k\Dot N)^2/N^2-k^2}{\sa k^2\,(k^2+2k\Dot N+4-\bar{\la})\,(k^2+k\Dot N)^2}\,,
\end{equation}
with $N=(2,\vec{0})$ and $\bar{\la}=\la/m_\mu^2$. Replacing $\la$ by $4m^2z$ and evaluating the $d^{\,4}k$ integral in the usual way results in the expression
\begin{equation}
E_{VP}=-\frac{m_\mu \al^6}{\pi^2}\,I(r)\,,
\end{equation}
with $r=m^2/m_\mu^2$ and 
\begin{equation}
I(r)=\int_r^\infty\!\frac{dz\,(2z+r)\,\sqrt{z-r}\,F(z)}{2z^{5/2}}\,.
\end{equation}
The function $F(z)$ is
\begin{equation}
F(z) = \frac{1}{6}\left[2-2\log(2)-4z\log(2)+(z-1)\log(z-1)-3z\log(z) +4z\sqrt{\frac{z}{z-1}} \log\left(\sqrt{z}+\sqrt{z-1}\right)\right].
\end{equation}
If the muon is in the loop, $r=1$ and Mathematica gives $I(1)=\pi^2/36$, leading to the usual result
\begin{equation}
\De E_{VP}^{\mu\bar{\mu}}=-0.27416\frac{m_\mu\,\al^6}{\pi^2}\,.
\end{equation}
For $r=m_e^2/m_\mu^2$, numerically integrating using NIntegrate gives
\begin{equation}
I(r)=1.0501-4.9129\,i\,,
\end{equation}
which results in an additional contribution
\begin{equation} \label{ee}
\De E_{VP}^{e\bar{e}}=-1.0501\frac{m_\mu\,\al^6}{\pi^2}\,.
\end{equation}

Combining Eq.\,(\ref{ee}) with the rescaled positronium result gives a frequency shift for the $\mu^+\mu^-$ atom of
\begin{equation}
\De\nu_{\mu^+\mu^-}^S=-285.18\;{\rm MHz}
\end{equation}
for the $\al^6$ singlet annihilation process.

There is a corresponding two-loop annihilation contribution to the triplet hyperfine interval \cite{Cung,Cung1,ABZ}, which can be rescaled to give
\begin{equation}
\De\nu_{\mu^+\mu^-}^T=-200.38\;{\rm MHz}\,.
\end{equation}
The net $\al^6$ annihilation contribution for the ground state hyperfine splitting is then
\begin{equation}
\De\nu_A(\al^6) = 84.80\;{\rm MHz}\,.
\end{equation}
This is quite small in comparison to the $\al^6\log(\al^{-1})$ ground state contribution \cite{ABZ}, which rescales to
\begin{equation}
\De\nu(\al^6\log(\al^{-1}))=\frac{5 m_\mu\al^6\log(\al^{-1})}{24}=3.95454\times 10^3\;{\rm MHz}\,.
\end{equation}

\end{document}